# Comparison of Gd addition effect on the superconducting properties of FeSe$_{0.5}$Te$_{0.5}$ bulks under ambient and high-pressure conditions

Manasa Manasa[1], Mohammad Azam[1], Tatiana Zajarniuk[2], Ryszard Diduszko[3], Jan Mizeracki[1], Tomasz Cetner[1], Andrzej Morawski[1], Andrzej Wiśniewski[2], Shiv J. Singh[1*]

[1]Institute of High Pressure Physics (IHPP), Polish Academy of Sciences, Sokołowska 29/37, 01-142 Warsaw, Poland

[2]Institute of Physics, Polish Academy of Sciences, aleja Lotników 32/46, 02-668 Warsaw, Poland

[3]Łukasiewicz Research Network Institute of Microelectronics and Photonics, Aleja Lotników 32/46, 02-668 Warsaw, Poland

*Corresponding author:

 Email: sjs@unipress.waw.pl

# Abstract

We have prepared a series of (FeSe$_{0.5}$Te$_{0.5}$ + $x$Gd) bulk samples, with $x$ = 0, 0.03, 0.05, 0.07, 0.1 and 0.2, through the convenient solid-state reaction method at ambient pressure (CSP). High gas pressure and high-temperature synthesis methods (HP-HTS) are also applied to grow the parent compound ($x$ = 0) and 5 wt% of Gd-added bulks. Structural, microstructural, transport and magnetic characterizations have been performed on these samples in order to draw the final conclusion. Our analysis results that the HP-HTS applied for the parent compound enhances the transition temperature ($T_c$) and the critical current density ($J_c$) with the improved sample density and intergrain connections. The lattice parameter '$c$' is increased with Gd additions, suggesting a small amount of Gd enters the tetragonal lattice of FeSe$_{0.5}$Te$_{0.5}$ and the Gd interstitial sites are along the $c$-axis. A systematic decrease of the onset transition temperature $T_c$ is observed with Gd additions, however, the calculated $J_c$ of these Gd-added samples is almost the same as that of the parent compound prepared by CSP. It specifies that there is no improvement of the grain connections or pinning properties due to these rare earth additions. However, Gd-added FeSe$_{0.5}$Te$_{0.5}$ bulks prepared by HP-HTS have revealed a slightly improved critical current density due to improved grain connections and sample density but have a lower transition temperature than that of the parent compounds.

# Introduction

The discovery of iron-based superconductors (FBS) [1] provides the second high $T_c$ superconductors after the first high $T_c$ cuprate superconductors [2]. Interestingly, these FBS are very rich in chemistry, and many kinds of doping have been reported [3, 4]. More than 100 compounds are available for this high $T_c$ material that can be categorized into six families based on the crystal structure of its parent compound [5, 6, 7], and the maximum transition temperature is reached up to 58 K [8]. In all these families, FeSe belonging to the 11 family has the simplest crystal structure and depicts the critical transition temperature ($T_c$) of 8 K at ambient pressure [9, 10, 11]. Edge-sharing tetrahedral $FeSe_4$ layers are the sole layers in FeSe, and they are stacked along the $c$-axis. No charge storage layer exists. A structural transition from tetragonal to orthorhombic occurs at about $T_s \sim 90$ K accompanied by the nematic phase [12]. However, various kinds of doping in 11 family have been reported, such as Co, Ni, Cu, Cr at Fe sites and S, Te at Se sites, to understand the chemical pressure effect [13, 10, 14], the superconducting mechanism, and to enhance the superconducting properties. In all these doping, the substitution of Te at Se sites has enhanced the $T_c$ value up to 15 K at ambient doping for 50% substitution [15, 16]. Furthermore, the applied pressure effect on FeSe has enhanced the transition temperature up to 36.7 K [15]. The reported upper critical field of 50 T and the critical current density of $\sim 10^4$ A/cm$^2$ (0 T, 5 K) have been obtained for the 11 family [15, 17], which also doesn't contain any harmful elements like arsenic. To enhance the superconducting properties of these materials, different kinds of methods have been applied, such as metal addition [18, 19, 20, 21], applied pressure studies, high-pressure synthesis [17], etc. by improving the grain boundaries [22].

One of the basic challenges of this 11 family is to prepare a completely pure superconducting phase due to the complicated phase diagram of FeSe [23, 10, 24, 11] which has many stable crystalline forms such as tetragonal $\beta$-$Fe_xSe$, hexagonal $\delta$-$Fe_xSe$, orthorhombic $FeSe_2$, monoclinic $Fe_3Se_4$, and hexagonal $Fe_7Se_8$, in which the tetragonal phase generally exhibits superconductivity with $T_c = 8$ K [11]. During the growth process, a number of these stable phases, particularly, hexagonal $\delta$-$Fe_xSe$ and hexagonal $Fe_7Se_8$, appear along with the primary tetragonal $\beta$-$Fe_xSe$ phase, but they lack the necessary superconducting properties. High-pressure synthesis of these materials is also not able to completely reduce the hexagonal phase but enhances the superconducting properties of Fe(Se,Te) [17]. To improve the superconducting properties of FBS, one of the most common methods at ambient pressure is the metal additions [6, 3, 18]. In this direction, many kinds of metal addition such as Pb, Sn,

Li, Ag, Potassium iodide (KI) [25, 26] *etc* have been reported [27, 18, 28, 29] and interestingly, these metal additions reduce the hexagonal phase of Fe(Se,Te) and promote the tetragonal phase formation which works well for a small amount of addition by enhancing the superconducting properties of 11 family [18].

In the case of $MgB_2$ superconductor [30], many transition metal additions have been reported, among which some unique properties have been observed with the rare earth (*RE*) additions. Many studies have been reported on the addition of *RE* elements and their compounds. Rare earth doping in $MgB_2$ such as Y, La, Dy, Ho, Nd *etc* has significantly enhanced the critical current density $J_c$ [30]. These *RE* react with boron (B) and form impurities phases such as $RE B_6$ and $RE B_4$ in $MgB_2$ superconductors because the larger diameter of *RE* ions makes it hard to introduce them into the Mg sites in the $MgB_2$ lattice. However, the enhancement of critical current density and upper critical have been substantially achieved in these alloyed compounds, whereas the critical transition temperature $T_c$ is almost unchanged. Similarly, various studies have also been reported for other superconductors, such as NbTi [31]. These studies motivated us to study and understand the superconducting properties of $FeSe_{0.5}Te_{0.5}$ bulks with rare earth additions. Additionally, there is no report based on the rare earth addition of Fe(Se,Te) bulks, which is the main motivation of this research paper.

Here, a series of Gd-added $FeSe_{0.5}Te_{0.5}$ bulks ($FeSe_{0.5}Te_{0.5} + x$Gd; $x = 0$, 0.03, 0.05, 0.07, 0.1, and 0.2) have been prepared by the solid-state reaction method at ambient pressure, *i.e.* convenient synthesis process (CSP) and by HP-HTS. These samples are well characterized by structural, microstructure, magnetic, and transport properties to understand the rare earth addition effects on the superconducting properties of the parent compound of $FeSe_{0.5}Te_{0.5}$. The detailed structural and microstructural analysis depicts that the hexagonal phase is reduced completely with Gd additions and is not observed again even at high amounts of Gd additions. HP-HTS of the parent compound ($x = 0$) are very effective for the enhancement of the superconducting phase, however, the HP-HTS growth of Gd-added samples reduces the transition temperature, and a slightly improved $J_c$ is observed due to the improved intergrain connections. Our various analyses suggest that Gd is entered the superconducting lattice and the hexagonal phase is completely reduced. However, the superconducting properties are decreased systematically with Gd additions.

# Experimental details

A series of Gd additions, FeSe$_{0.5}$Te$_{0.5}$ (FeSe$_{0.5}$Te$_{0.5}$ + $x$Gd), has been prepared by CSP at ambient pressure with $x$ = 0, 0.03, 0.05, 0.07, 0.1, and 0.2. Basically, these samples followed a two-step process. In the first step, the starting precursors: Fe powder (99.99% purity, Alfa Aesar), Se (99.99% purity, Alfa Aesar), and Te (99.99% purity, Alfa Aesar) were mixed according to stoichiometric FeSe$_{0.5}$Te$_{0.5}$ composition in an agate mortar and then prepared into the pellets of 12 mm diameter. These pellets were sealed into an evacuated quartz tube, which was placed inside the furnace and heated at 600°C for 11 hours. After this step, the quartz tube was opened inside the glove box and the obtained pellets were reground. Then, we mixed Gd metal (99.9%, Alfa Aesar) with 0 wt% ($x$ = 0), 3 wt% ($x$ = 0.03), 5 wt% ($x$ = 0.05), 7 wt% ($x$ = 0.07), 10 wt% ($x$ = 0.10), and 20 wt% ($x$ = 0.2), and each Gd added samples were prepared at a weight of 1 gram. After the grinding process, this powder was prepared in a pellet of 12 mm diameter and sealed inside an evacuated quartz tube. Again, this prepared ampoule was heated at 600°C for 4 h inside the furnace. The final prepared pellet has a diameter of 12 mm and a 2.5 mm thickness. All synthesis processes were performed inside a high-purity argon-filled glove box. Various batches of these samples were produced under the same synthesis conditions to confirm their reproducibility. More details about the synthesis process is reported elsewhere [18, 27] and the details about the prepared bulks are listed in Table 1.

We have also prepared Gd added FeSe$_{0.5}$Te$_{0.5}$ bulks with the HP-HTS method, which can apply the inert gas pressure of up to 1.8 GPa inside a cylindrical chamber. This high-pressure chamber is equipped with a single zone or three zone furnace. More details are discussed elsewhere [17]. In the first step, the initial precursors Fe, Se, and Te were mixed according to the stoichiometric composition and heated in a furnace at 600°C for 11 h, as discussed above as the first step. In the second step, Gd metal is added to the prepared FeSe$_{0.5}$Te$_{0.5}$ after the first step and sealed inside a Ta-tube under an Ar-gas atmosphere through an ARC melter. This sealed metal tube is placed inside the pressure chamber and applied to a pressure of 500 MPa at 600°C for 1 hour through HP-HTS. In a previous investigation [17], we used HP-HTS to optimize the development of bulk Fe(Se,Te), and the high superconducting characteristics could be attained by synthesizing the material at 600 °C for an hour at 500 MPa. Table 1 includes a list of the prepared samples.

For the structural characterization of these samples, we have measured X-ray diffraction patterns using Rigaku SmartLab 3kW diffractometer with filtered *Cu-Kα* radiation (wavelength: 1.5418 Å, power: 30 mA, 40 kV), and a Dtex250 linear detector. A slow scan of the measurement profile from 5° to 70° with a very small step of 0.01°/min was used to measure our samples. Furthermore, ICDD PDF4+ 2021 standard diffraction patterns database and Rigaku's PDXL software were applied to perform the profile analysis. On the basis of these analyses, the quantitative values of impurity phases (%), and lattice parameters were calculated for various samples. Zeiss Ultra Plus field-emission scanning electron microscope equipped with the EDS microanalysis system by Bruker mod. Quantax 400 with an ultra-fast detector was carried out for the detailed macrostructural analysis and in the mapping of the constituent elements. Quantum Design PPMS attached to a vibrating sample magnetometer (VSM) was used to characterize the magnetic properties of these materials in the temperature range of 5-25 K and in the magnetic field up to 9 T. The magnetic susceptibility was measured in zero-field cooled (ZFC) and field-cooled modes at an applied magnetic field of 50 Oe with a slow temperature scan. The variation of resistivity with temperature was measured by a closed-cycle refrigerator (CCR) in a zero magnetic field with various current (5, 10, 20 mA), where all data were collected with a very slow warming process.

## Results and discussion

The collected data of the powder X-ray diffraction patterns for various Gd-added $FeSe_{0.5}Te_{0.5}$ ($FeSe_{0.5}Te_{0.5} + x$Gd) samples are illustrated in Figure 1(a). All samples have the main tetragonal phase with the space group *P4/nmm* as similar to the parent compound $FeSe_{0.5}Te_{0.5}$ ($x = 0$) [10]. The parent compound also has around 5-6% hexagonal phase as an impurity phase, which is well in agreement with the previous reports for 11 family [32, 10, 33]. Interestingly, a very small amount of Gd addition, such as 3% weight has reduced the hexagonal phase significantly, but a very small amount of Gd metal is also observed as a secondary phase. Furthermore, 5 wt% of Gd added sample shows almost no hexagonal phase and also no other impurity phase is observed. With the further addition of Gd, the hexagonal phase is completely suppressed and not observed again, however, the amount of the Gd metal as an impurity is increased with higher Gd additions, as listed in Table 2. Since Gd metal is very sensitive to air, so the sample having a high amount of Gd has also shown $Gd_2O_3$ impurity, which could be possible during the transfer of the powder XRD samples from the glove box. Since the sample with $x = 0.05$ has an almost clean pristine phase, we decided to grow the parent compound ($x = 0$) and 5 wt% of Gd added bulks ($x = 0.05$) by HP-HTS to understand the high pressure synthesis effects. In Figure 1(a), the parent sample ($x = 0\_HIP$) and 5% weight Gd added sample ($x = 0.05\_HIP$) prepared by HP-HTS are also shown. Interestingly, the parent compound, *i.e.*, $x = 0\_HIP$, has almost the same amount of hexagonal phase as $x = 0$ sample obtained by CSP. Whereas the sample $x = 0.05$ prepared through HP-HTS have nearly no hexagonal phase, a very small amount of GdSe phase is observed, as listed in Table 2. It suggests that the HP-HTS method somehow supports the formation of the GdTe/GdSe phase instead of the hexagonal phase under high-pressure growth of Gd added $FeSe_{0.5}Te_{0.5}$. Fitted XRD data for $x = 0$, 0.05, and 0.05_HIP are shown in Figures 1(b), 1(c), and 1(d), where the experimentally observed and calculated intensity, and the difference between these two curves are depicted. The refinement is well fitted with a tetragonal phase with space group *P4/nmm*. As listed in Table 2, the lattice parameter of the parent compound prepared through CSP and HP-HTS has the lattice parameter ($a = 3.7950$ Å, $c = 5.9713$ Å) and ($a = 3.7976$ Å, $c = 5.9679$ Å) at ambient and applied pressure, respectively, as previously reported for polycrystalline samples ($a = 3.7909$ Å, $c = 5.9571$ Å) [11] and single crystals ($a = 3.815$Å, $c = 6.069$ Å) of $FeSe_{0.5}Te_{0.5}$ [34, 35, 36]. The lattice parameters and the qualitative values of the impurity phases of all samples are listed in Table 2. Interestingly, the hexagonal phase is reduced completely even with a small addition of Gd, which is similar to other metal additions such as Pb [27], Sn [18], Li [28], Ag [37], etc., but this hexagonal phase could not be reduced by the high-pressure synthesis method [17]. Data in Table 2 indicates the slight enhancement of lattice parameter '$c$' with Gd additions with respect to the parent compound ($x = 0$), which suggests that Gd metal was somehow entered inside the tetragonal lattice of superconducting $FeSe_{0.5}Te_{0.5}$ phase and Gd interstitial sites are along the *c*-axis, but it doesn't change the amount of Se or Te concentrations. The effect of Gd addition is comparable to that of the earlier study based on Li doping in Fe(Se,Te) [28], where Li entered the superconducting lattice. Due to the presence of various impurity phases, there could be a slightly higher error in the refinement, especially for larger amounts of Gd additions. One important point is that, like other metal addition, Gd-addition promotes the formation of the tetragonal phase by reducing the hexagonal phase. However, Gd metal also manifests as an impurity phase, as listed in Table 2.

To analyse the actual composition of constituent elements and their distribution inside the sample, we have performed the elemental mapping and the energy dispersive X-ray (EDAX) which is shown in Figure 2. Data in Table 3 indicate the actual concentration of various elements from the EDAX analysis. The parent compound ($x = 0$) shows the homogeneous distribution of all elements, whereas 5% weight Gd addition (Figure 2(ii)) has the better homogeneous distribution of these elements compared to the parent compound (Figure 2(i)), which could be due to the reduced hexagonal phase. In very few places, we have observed Gd as a metal form. The inhomogeneity of these constituent elements is increased with the further addition of Gd. Figure 2(iii) shows the elemental mapping for the samples with $x = 0.20$, where Gd metal is observed in many places with large areas, as discussed with XRD analysis. As shown in Figure 2, the parent compound exhibits a molar ratio of 1:0.49:0.51 (Fe:Se:Te), which is nearly identical to the bulks $FeSe_{0.5}Te_{0.5}$ with a small amount of Gd additions. As mentioned in Table 3, the nominal weight of the Gd addition is nearly equal to the actual concentration, but with a high number of Gd additions, the deviation in the molar ratio is slightly increased in comparison to that of the parent compound. These results indicate that Fe, Se, and Te concentrations are not affected by Gd additions, but non-uniform Gd distributions are clearly visible at high Gd addition levels.

Figures 3 (i) and 3(ii) show the mapping for the parent compound ($x = 0\_HIP$) and Gd-added sample ($x = 0.05\_HIP$) prepared using the HP-HTS technique. Figure 3(i) shows that the homogeneity of the various elements for $x = 0\_HIP$ is the same as that of the parent compound prepared in ambient conditions (Figure 2(i)). Fe, Se, and Te are distributed relatively uniformly across the sample $x = 0.05\_HIP$, but some regions are rich in Gd and Se elements, which points to the formation of the GdSe phase as proposed by the XRD study. This sample ($x = 0.05$) prepared at ambient pressure has a somewhat different mapping from this Gd-added sample ($x = 0.05\_HIP$) (Figure 2(ii)). It implies that the homogeneity of the Gd and Se element distributions is decreased by the high-pressure effect. As shown in Table 3 and also by XRD analysis and mapping, HP-HTS for the Gd-added sample ($x = 0.05\_HIP$) reduced the actual Se-content from the stoichiometry of $FeSe_{0.5}Te_{0.5}$ and produced an addition phase, GdSe. For these reasons, we have not prepared any other Gd added $FeSe_{0.5}Te_{0.5}$ bulks using HP-HTS.

In order to understand the microstructural analysis, the polycrystalline samples were polished manually using various sandpapers with different grades inside the glove box to collect backscattered scanning electron microscopy (BSE-SEM, revealing chemical contrast) images at different magnifications. Figure 4 depicts low to high-magnification images for three samples with $x = 0$, 0.05, and 0.2. In these images, light gray, white, and black contrasts are observed corresponding to the phases of $FeSe_{0.5}Te_{0.5}$, $Gd_2O_3$, and pores, respectively. Figure 4(a)–(c) demonstrates that the sample with $x = 0$ has two contrasts: light gray and black contrasts, and on the microscale, the microstructural images are nearly homogeneous. Numerous micropores as well as many well-connected, disk-shaped grains are observed. As illustrated in Figure 4(d)-(f), a minor addition of Gd ($x = 0.05$) tends to result in larger grains and slightly larger pores. Overall, the microstructural of the sample with $x = 0.05$ is almost identical to the parent compound. However, we have not observed Gd metal in these BSE images for $x = 0.05$, as previously noted via XRD and mapping. With further increasing of Gd additions, it seems that grain size and pore size are increased. At higher amount of Gd additions, we have observed, the white contrasts related to $Gd_2O_3$ which confirmed the presence of Gd metals as an impurity phase, as similar to XRD analysis. Figures 4(h)-(j) depict the BSE images for the sample with $x = 0.20$, where a larger grain is observed with the enhanced pore size. As shown in Figure 4, white contrast ($Gd_2O_3$) is seen in more regions and bigger sections of the sample for Gd additions ($x > 0.05$), and the increased pore size is also noticeable. Because of the increased impurity phase ($Gd_2O_3$) between $FeSe_{0.5}Te_{0.5}$ grains, grain-to-grain connections are often drastically reduced, and intergranular supercurrent pathways are severely blocked. These investigations indicate that, in comparison to bulk samples with $x = 0$, Gd addition lowers the grain connections due to larger grain and pore sizes. As a result, it implies that neither grain connectivity nor material density have improved.

Figures 5 show BSE images for the parent compound ($x = 0$) and Gd-added sample ($x = 0.05\_HIP$) prepared through HP-HTS techniques. Compared to the samples prepared through CSP (Figure 4), these samples are more compact. The sample $x = 0\_HIP$ has several well-connected grains which reduces the size of the pores. The sample $x = 0.05\_HIP$ also has better grain connections compared to the sample $x = 0.05$, but $Gd_2O_3$ or GdSe is also observed as an impurity phase, which weakens the grain connections. The most prominent phase of $Gd_2O_3$ or GdSe is randomly observed as a white contrast in the bulk sample at many places, such as inside grains and at grain boundaries, but the size of pores as a black contrast is smaller than in the samples with $x = 0.05$. These pores and impurity phases are what cause the samples' weak grain connections, and Figures 5(d)–(f) appear to show that the observed grains are disc-shaped. As reported for other iron-based superconductors [38, 39], we have not found any microcracks in our samples at the grain borders or within the grains. The sample density is found to be around 51%, 50%, 48%, 72%, and 56%, respectively, for $x = 0$, 0.05, 0.07, 0.0\_HIP, and 0.05\_HIP, based on the theoretical density of Fe(Se,Te) (6.99 g/cm$^3$) [11]. The small amount of Gd added sample, $i.e.$, $x = 0.05$, has almost the same density as that of the parent compound ($x = 0$). However, the HP-HTS method has slightly improved the density for the sample $x = 0.05\_HIP$, whereas a large improvement is observed for the parent compound $x = 0\_HIP$, as also demonstrated by the microstructural analysis. The microstructural analysis in Figures 4 and 5 reveals that the addition of Gd improves the grain size and increases the pore size, but that the sample density is almost identical to that of the parent compound ($x = 0$) regardless of whether samples are prepared using CSP or HP-HTS. A large amount of Gd additions degrades both the phase purity and the cleanliness of grain boundaries with large pores. It is well known that non-superconducting phases at the grain boundaries create an obstacle to the superconducting properties, as reported for 1111 [38, 4], 122 [40], and also for the 11 family [41] [27]. As a result, our analysis suggests that even a very small amount of the rare earth addition to FeSe$_{0.5}$Te$_{0.5}$ is ineffective at boosting material density or enhancing grain size and connectivity.

To confirm the Meissner effect of these samples, the measured DC magnetic susceptibility ($\chi = 4\pi M/H$) is shown in Figure 6(a)-(b) for samples $x = 0$ and $x = 0.05$, $x = 0.0\_HIP$, and $x = 0.05\_HIP$ measured under an applied magnetic field of 50 Oe in the temperature range 5-20 K for zero-field-cooled (ZFC) and field-cooled (FC) magnetization curves. The normalized magnetic susceptibility was calculated and depicted in Figure 6(a)-(b) for a comparison, which confirms a bulk superconductivity for these samples. The parent compound ($x = 0$) shows a superconducting transition of 14 K and, 5% Gd added FeSe$_{0.5}$Te$_{0.5}$ ($x = 0.05$) has almost the same $T_c$ as that of the parent compound. Also, ZFC and FC behaviour of these two samples is very similar, as depicted in Figure 6(a), which confirms that there is no change in Se or Te concentration inside the stoichiometry of FeSe$_{0.5}$Te$_{0.5}$, as also indicated by structural and microstructural analysis. Figure 6(b) shows the normalized magnetization behaviour for the bulk samples prepared by the HP-HTS method. The sample $x = 0\_HIP$ has an enhanced onset transition temperature by 1.2 K with a sharper transition, suggesting a good grain connection. The sample $x = 0.05\_HIP$ has a broader transition due to the presence of the impurity phase of GdSe but the onset transition is almost the same as that of $x = 0$. All bulk samples have depicted the single-step transition, suggesting the intergranular properties of these bulks are comparable to those reported for other families of FBS [42]. Therefore, these analyses also support the conclusion that Gd addition does not affect the superconducting transition of FeSe$_{0.5}$Te$_{0.5}$ whether it was prepared through CSP or HP-HTS, which is similar to what was found by microstructural analysis and XRD measurements. The almost same $T_c$ value of the Gd added bulks prepared by CSP confirms that Gd does not change in Te/Se concentrations.

The variation of resistivity ($\rho$) with respect to the temperature is illustrated in Figure 7(a)-(c) for the nominal compositions of polycrystalline FeSe$_{0.5}$Te$_{0.5}$ + $x$Gd ($x = 0$–0.2) in a zero magnetic field prepared by CSP. Due to the structural phase transition, the parent FeSe$_{0.5}$Te$_{0.5}$ ($x = 0$) exhibits a large anomaly in resistivity at a temperature below ~110 K [16, 33]. The small amount of Gd addition, *i.e.*, $x = 0.03$, has enhanced the normal state resistivity, which might be possible due to the inhomogeneous distribution of a small amount of Gd contents. However, the behaviour of the resistivity curve is very similar to that of the parent compound. With the further increase of Gd additions, the resistivity is reduced in a systematic way due to the increased concentration of Gd metallic element. Interestingly, the overall resistivity behaviour of all Gd-added samples up to 10 weight% ($x = 0.10$) has almost similar behaviour to the parent compound. The sample $x = 0.2$ has a completely different behaviour than the other samples, *i.e.*, metal to insulation behaviour, as shown in the inset of Figure 7(a). As discussed with XRD and microstructural analysis, this sample has the impurity phase of $Gd_2O_3$, and this could be a reason for this metal to insulator behaviour. The low-temperature behaviours of these samples are depicted in Figure 7(b) in the temperature range from 5 K to 20 K, where all samples exhibit the superconducting transition. The parent compound displays a superconducting transition of 14.8 K but has a slightly broader transition. Interestingly, the low resistivity behaviour of these Gd-added samples is not very systematic. The sample with $x = 0.03$ has an onset $T_c$ of 14 K and an offset $T_c$ of 11.8 K, *i.e.*, transition width of 2.2 K. Further increase of Gd addition has marginally improved the transition width of 1.9 K for $x = 0.05$. The sample has an onset transition of 14.3 K and an offset transition of 12.4 K. In the next step, the sample with $x = 0.07$ has a $T_c$ value of 13.1 K with an offset $T_c$ of 9.81 K and a transition width (~3.3 K) almost similar to the parent compound. Interestingly, slightly higher Gd additions, *i.e.*, $x = 0.1$, show an onset $T_c$ of 13.5 K and an offset $T_c$ of 11.9 K with a transition width of 1.6 K. A large amount of Gd addition, *i.e.*, $x = 0.2$, has an onset transition of 12.7 K, but no zero resistivity is observed up to 7 K. It suggests that there is a non-superconducting phase inside the bulks, which is also confirmed by the XRD and microstructural analysis.

To comprehend the grain connections of these samples prepared through CSP, we have measured the temperature dependence of the resistivity under various currents ($I = 5$, 10, and 20 mA). The individual grain effect, *i.e.*, the intragrain effect, is related to the onset transition temperature ($T_c^{onset}$), whereas the grain connections, *i.e.*, the intergrain effect, represent the offset transition temperature ($T_c^{offset}$) [43, 44]. These effects can be understood by the resistivity measurements under different applied currents. Figure 7(c) illustrates the resistivity behaviour in the low temperature region for these samples with three different currents $I = 5$, 10, and 20 mA to investigate the grains and grain connectivity behaviours. The bulk sample with $x = 0.03$ has a broader transition with various currents where the onset and offset transitions are both sensitive with various currents, as shown in Figure 7(c). However, the samples with $x = 0.05$ and 0.07 have almost no transition broadening, which suggests good grain connections. The bulk samples with $x = 0.1$ have slightly broader transition and seem very similar to the parent compound, which could be due to the enhanced impurity phase of Gd metal as seen in the XRD patterns. These analyses suggest that 5 to 7% weight of Gd addition has slightly improved the grain connections compared to the parent compound ($x = 0$) and other Gd-added samples. These findings corroborate the study of microstructural investigations that was previously described. Compared to our results with previous studies based on metal additions such as Sn and Pb added samples [18], the mid-range addition of Gd additions to $FeSe_{0.5}Te_{0.5}$ has almost no broadening of the transition with respect to the applied current, which suggests better grain connectivity but a lower superconducting transition due to the inhomogeneity of the bulks. Interestingly, Gd addition has a similar impact on the onset and offset of superconducting transitions. These results are well in agreement agreed with microstructural and XRD analyses.

Figures 7(d) and 7(e) show the temperature dependence of resistivity behaviour for the $FeSe_{0.5}Te_{0.5}$ bulk prepared by the HP-HTS process for $x = 0$_HIP and 0.05_HIP. In comparison to CSP, the resistivity of the parent compound ($x = 0$) has been reduced by almost 50% through HP-HTS, which might be due to improved sample density and proper grain connections, as discussed with the microstructural analysis [17]. In the case of 5 wt% of Gd addition, HP-HTS processes are not very effective due to the formation of the GdSe phase. Due to this impurity phase, the resistivity depicts the metal-to-insulation behaviour, which is almost identical to the sample $x = 0.2$. Figure 7(e) depicts the low-temperature behaviour of these samples: $x = 0$, 0_HIP, and 0.05_HIP. The sample $x = 0$_HIP shows an enhanced transition temperature of 1.2 K, as also confirmed by the magnetization study, and the transition width is improved with respect to the sample $x = 0$. However, 5 wt% of Gd-added sample prepared by HP-HTS has an onset transition of 11.8 K, but the zero resistivity is not reached by the measured temperature range up to 7 K (Figure 7(e)). This behaviour, along with XRD and microstructural analysis, supports the existence of a non-superconducting phase and a decrease in Se contents in this sample.

To figure out the critical current density $J_c$, the magnetic hysteresis loops $M(H)$ at a constant temperature of 7 K were performed with the rectangular-shaped sample $x = 0$, 0.05, and 0_HIP, 0.05_HIP. Similar to the previous reports for the parent $Fe(Se,Te)$ compounds [27, 45, 46], these samples depict the magnetic hysteresis loops under ferromagnetic effects due to the presence of a very tiny amount of iron, which is non-observable for the XRD measurements. The $M(H)$ loop for a sample with $x = 0$_HIP is shown as an inset of Figure 8 after the subtraction of the normal state magnetization, *i.e.*, $M(H)$ loop at 22 K. In a similar way, Gd-added samples have also depicted the magnetic hysteresis loop with a large background. These measured *M-*

*H* loops enable for the calculation of the critical current properties of these samples, which is an important parameter from a practical point of view. The Bean model [47], which is popular and reliable for iron-based superconductors, is one of many models [48, 47] that have been proposed to obtain the $J_c$ values. We have performed the $J_c$ calculation by using the formula $J_c = 20\Delta m/Va(1-a/3b)$ [47], where $\Delta m$ is the hysteresis loop width, *V* is the volume of the sample, and shorter and longer edges of the sample are *a* and *b*, respectively. The magnetic field dependence of the critical current density $J_c$ for these samples is shown in Figure 8 at 7 K and for the measured magnetic field up to 9 T. The parent compound *x* = 0 has shown the maximum $J_c$ values of the order of $10^2$ A/cm$^2$ at 0 T which is almost similar to the previously reported paper on the basis of the CSP method [27]. The high-pressure synthesis of this sample, *i.e.*, *x* = 0_HIP, has shown the almost two orders of magnitude enhancement of $J_c$ of the parent compound (*x* = 0), which is depicted in Figure 8. This $J_c$ value of the order of $10^4$ A/cm$^2$ is the highest value of FeSe$_{0.5}$Te$_{0.5}$, as reported by the melted synthesis route [49] or other methods [17]. Interestingly, 5 wt% of Gd added FeSe$_{0.5}$Te$_{0.5}$ (*x* = 0.05) has the same $J_c$ value and almost similar behaviour to the parent compound (*x* = 0), which could mean that this sample (*x* = 0.05) has the same sample density and the same microstructural likeness to the parent compound. High-pressure synthesis of the Gd added sample, *i.e.*, *x* = 0.05_HIP, has slightly improved $J_c$ values than those of the parent compounds (*x* = 0) in the whole magnetic field range up to 9 T. The calculated $J_c$ of all samples has similar behaviour. This $J_c$ improvement of the sample *x* = 0_HIP and 0.05_HIP might be due to improved grain connection and the improved sample density by the high-pressure growth method, as discussed above with microstructural analysis, which is capable of providing effective flux pinning centres. The same observation has also been also reported for MgB$_2$ [37] and NbTi- based superconductors [31], where Ag or Gd addition enhances $J_c$ values due to extra pinning centres. One should note that 5 wt% of Gd-added bulks (*x* = 0.05) have almost the same $J_c$ values [27], whereas Sn and Pb additions enhance the $J_c$ values by one order of magnitudes by improving the intergranular current compared to other metal additions [30]. Sn metal additions also work well to improve and enhance the $J_c$ values for Sn-added SmFeAs(O,F) [43]. These analyses suggest that Sn, Ag, or Pb, or cometal addition [20, 18] can be the most effective metal to enhance the $J_c$ value for FeSe$_{0.5}$Te$_{0.5}$ samples compared to the rare earth (Gd) addition. By using the calculated $J_c$ values at 7 K, vortex pinning force density, $F_p$ has been calculated by the formula $F_p = \mu_0 H \times J_c$ [50]. Similar to the previous studies [32, 46], the maximum $F_p$ values of (~0.3-0.4 GN/m$^3$) of the polycrystalline sample with *x* = 0 and *x* = 0.05 are nearly identical as those reported values (~0.1-1 GN/m³) for Fe(Se,Te) bulks. The sample $x = 0.05\_HIP$ has slightly higher $F_p$ values (~0.6 GN/m³) than those of other samples ($x = 0$ and $0.05$); which is in nice agreement with the $J_c$ enhancement, as depicted in Figure 8. Curiously, the parent samples ($x = 0\_HIP$) by HP-HTS have a significant increase in the maximum $F_p$ value (~15 GN/cm³), supporting the high $J_c$ values and improvement of pinning centres for this sample. These studies confirm that Gd additions are not improving the appropriate pinning centres in the direction of the enhancement of critical current properties, which can be concluded from the previous reports for Ag-added MgB$_2$ [37] and Sn-added other FBS bulk samples [43]. Even, high-pressure growth and high-pressure sintering are also not helpful for Gd-added Fe(Se,Te) bulks to further improve the $J_c$ and $F_p$ values.

To reach our finding based on our study, Figure 9 shows the variation of the lattice parameter 'c', transition temperature $T_c^{onset}$, the transition width ($\Delta T$), room temperature resistivity ($\rho_{300K}$), and $RRR$ ($\rho_{300K}/\rho_{20K}$) with respect to the weight concentration of Gd added FeSe$_{0.5}$Te$_{0.5}$ samples ($x$). Interestingly, the lattice parameter 'c' is slightly increased with Gd addition, as shown in Figure 9(a), which suggests that a small amount of Gd enters the superconducting lattice either at Se sites or Te sites, however, some more detailed studies are needed in this direction. Even the HP-HTS process also enhanced the lattice 'c' for sample $x = 0.05\_HIP$ which could be possible due to the reduced amount of Te/Se from the FeSe$_{0.5}$Te$_{0.5}$ composition because this sample had GdSe as an impurity phase, as confirmed from XRD measurements. The $T_c^{onset}$ is reduced with Gd addition in a systematic way, which suggests that Gd enters inside the tetragonal lattice and reduces the superconducting transition, as depicted in Figure 9(b). The parent compound through HP-HTS has enhanced the $T_c$ around 1.2 K but 5 wt% Gd addition ($x = 0.05\_HIP$) reduces the onset $T_c$ due to the decreased Se-concentration. Figure 9(c) illustrates that the value of transition width $\Delta T$ ($= T_c^{onset} - T_c^{offset}$) is also suppressed with Gd addition and suggests the inhomogeneity of the sample which is supported by XRD and microstructural analysis. The room temperature resistivity ($\rho_{300K}$) is shown in Figure 9(d), and interestingly, is reduced with Gd addition due to the metallic nature of Gd. The sample with $x = 0.05\_HIP$ has a higher resistivity than the sample with $x = 0.0\_HIP$ due to the presence of GdSe phase. The residual resistivity ratio ($RRR$) of these samples is depicted in Figure 9(e), which decreased with Gd addition, even for the sample with $x = 0.05\_HIP$, which suggests that homogeneity, grain connections, and phase purity of the samples are reduced with Gd addition prepared either through CSP or HP-HTS. All of the studies clearly specify that Gd addition is not an effective way to improve the superconducting properties of the 11 family. On the other hand, adding a very small amount of Pb and Sn acts as an effective pinning centre [18] which increases the critical current density compared to the parent compound. This analysis suggests that rare earth Gd addition does not work as an effective method to improve both the superconducting properties and the intergranular properties.

## Conclusion

The superconducting properties of $FeSe_{0.5}Te_{0.5}$ bulks are explored with the rare earth Gd additions, for the first time, by the synthesis of a series of samples through CSP and HP-HTS. Structural analysis suggests that a small amount of Gd-addition entered the superconducting tetragonal lattice of $FeSe_{0.5}Te_{0.5}$, and due to this, the lattice parameter '$c$' is increased systematically with these additions. Although the sample density of Gd added bulk is nearly identical to that of the parent compound whereas grain size, and pore size are increased with Gd addition, as proposed by the microstructural analysis. Furthermore, the superconducting transition $T_c$ of Gd added $FeSe_{0.5}Te_{0.5}$ bulks prepared either though CSP or HP-HTS is reduced with Gd additions, whereas $J_c$ is almost the same as that of the parent compound through CSP but slightly improved by HP-HTS in the measured magnetic field up to 9 T due to the slightly enhanced sample density. Critical current analysis suggests that Gd addition is not suitable for providing additional pinning centres. HP-HTS processed $FeSe_{0.5}Te_{0.5}$ bulks ($x = 0\_HIP$) has improved the $T_c$ value by 1.2 K and almost two order of magnitude of the $J_c$ value. Our studies based on 11 family confirm that rare earth (Gd) additions cannot be a potential way to improve the intergrain connections, sample qualities, or superconducting properties of iron-based superconductors.

**Acknowledgments:**

The work was supported by SONATA-BIS 11 project (Registration number: 2021/42/E/ST5/00262) funded by National Science Centre (NCN), Poland. SJS acknowledges financial support from National Science Centre (NCN), Poland through research Project number: 2021/42/E/ST5/00262.

**Table 1:**

Details about the prepared FeSe$_{0.5}$Te$_{0.5}$ + $x$Gd samples and their synthesis conditions. CSP is used for "Conventional synthesis method at ambient pressure" and HP-HTS is used for "High gas pressure and high temperature synthesis method".

| Weight% of Gd addition | Sample code | Synthesis conditions | Growth method |
|---|---|---|---|
| 0 | $x = 0$ | *First step:* heated at 600 °C, 11 h, 0 MPa ↓ *Second step:* heated at 600 °C, 4 h, 0 MPa | CSP |
| 3% | $x = 0.03$ | *First step:* heated at 600 °C, 11 h, 0 MPa ↓ *Second step:* Gd addition and heated at 600 °C, 4 h, 0 MPa | CSP |
| 5% | $x = 0.05$ | *First step:* heated at 600 °C, 11 h, 0 MPa ↓ *Second step:* Gd addition and heated at 600 °C, 4 h, , 0 MPa | CSP |
| 7% | $x = 0.07$ | *First step:* heated at 600 °C, 11 h, 0 MPa ↓ *Second step:* Gd addition and heated at 600 °C, 4 h, 0 MPa | CSP |
| 10% | $x = 0.1$ | *First step:* heated at 600 °C, 11 h, 0 MPa ↓ *Second step:* Gd addition and heated at 600 °C, 4 h, 0 MPa | CSP |
| 20% | $x = 0.20$ | *First step:* heated at 600 °C, 11 h, 0 MPa ↓ *Second step:* Gd addition and heated at 600 °C, 4 h, 0 MPa | CSP |
| 0 | $x = 0\_HIP$ | *First step:* heated at 600 °C, 11 h, 0 MPa ↓ *Second step:* heated at 600 °C, 1 h, 500 MPa | HP-HTS |
| 5% | $x = 0.05\_HIP$ | *First step:* heated at 600 °C, 11 h, 0 MPa ↓ *Second step:* Gd addition and heated at 600 °C, 1 h, 500 MPa | HP-HTS |

**Table 2:**

A list of the calculated lattice parameters 'a' and 'c', and the impurity phases for $FeSe_{0.5}Te_{0.5} + x$Gd samples is provided. Rigaku's PDXL software and the ICDD PDF4+ 2021 standard diffraction patterns database have been used for the quantitative analysis of impurity phases (%) through the refinement of the measured XRD data.

| Sample (*x*) | Lattice 'a' (Å) | Latice 'c' (Å) | Hexagonal phase (%) | Gd phase (%) | $Gd_2O_3$ (%) | GdSe (%) |
|---|---|---|---|---|---|---|
| 0 | 3.7950(1) | 5.9713(3) | ~5 | - | - | - |
| 0.03 | 3.7989(2) | 5.9675(3) | ~2 | - | - | - |
| 0.05 | 3.7983(8) | 5.9744(2) | ~1 | - | - | - |
| 0.07 | 3.7993(6) | 5.9863(3) | - | ~1-2 | - | - |
| 0.10 | 3.7993(9) | 5.9892(4) | - | ~2-3 | - | - |
| 0.20 | 3.7977(3) | 5.9927(5) | - | ~3 | ~2 | - |
| 0_HIP | 3.7976(6) | 5.9679(1) | ~6 | - | - | - |
| 0.05_HIP | 3.8047(2) | 6.0463(4) | - | - | - | ~3 |

**Table 3:**

List of the molar ratio of various elements presented in $FeSe_{0.5}Te_{0.5} + x$Gd bulks.

| Sample ($x$) | Fe Molar Ratio | Se Molar Ratio | Te Molar Ratio | Gd Molar Ratio |
|---|---|---|---|---|
| 0 | 1 | 0.49 | 0.5 | - |
| 0.03 | 1 | 0.48 | 0.52 | 0.036 |
| 0.05 | 1 | 0.48 | 0.49 | 0.06 |
| 0.07 | 1 | 0.47 | 0.49 | 0.069 |
| 0.1 | 1 | 0.49 | 0.5 | 0.1 |
| 0.2 | 1 | 0.52 | 0.48 | 0.20 |
| 0_HIP | 1 | 0.50 | 0.51 | - |
| 0.05_HIP | 1 | 0.43 | 0.50 | 0.045 |

**Figure 1:** **(a)** X-ray diffraction patters (XRD) of powdered FeSe$_{0.5}$Te$_{0.5}$ + $x$Gd ($x$ = 0, 0.03, 0.05, 0.07, 0.1 and 0.2, 0_HIP, 0.05_HIP) samples. The fitted XRD patterns with the experimental, calculated diffraction patterns and their differences at the room temperature are shown for the sample with **(b)** $x$ = 0, **(c)** $x$ = 0.05 **(d)** $x$ = 0.05_HIP. Instead of the nominal composition of FeSe$_{0.5}$Te$_{0.5}$, the tetragonal phase of Fe$_{1.1}$Se$_{0.5}$Te$_{0.5}$ was observed as the real composition of the superconducting phase. One hexagonal phase, Fe$_7$Se$_8$, was has been found and is depicted as 'H' in figure (a). Table 1 contains a list of the obtained phases as well as the lattice parameters "$a$" and "$c$" that were obtained.

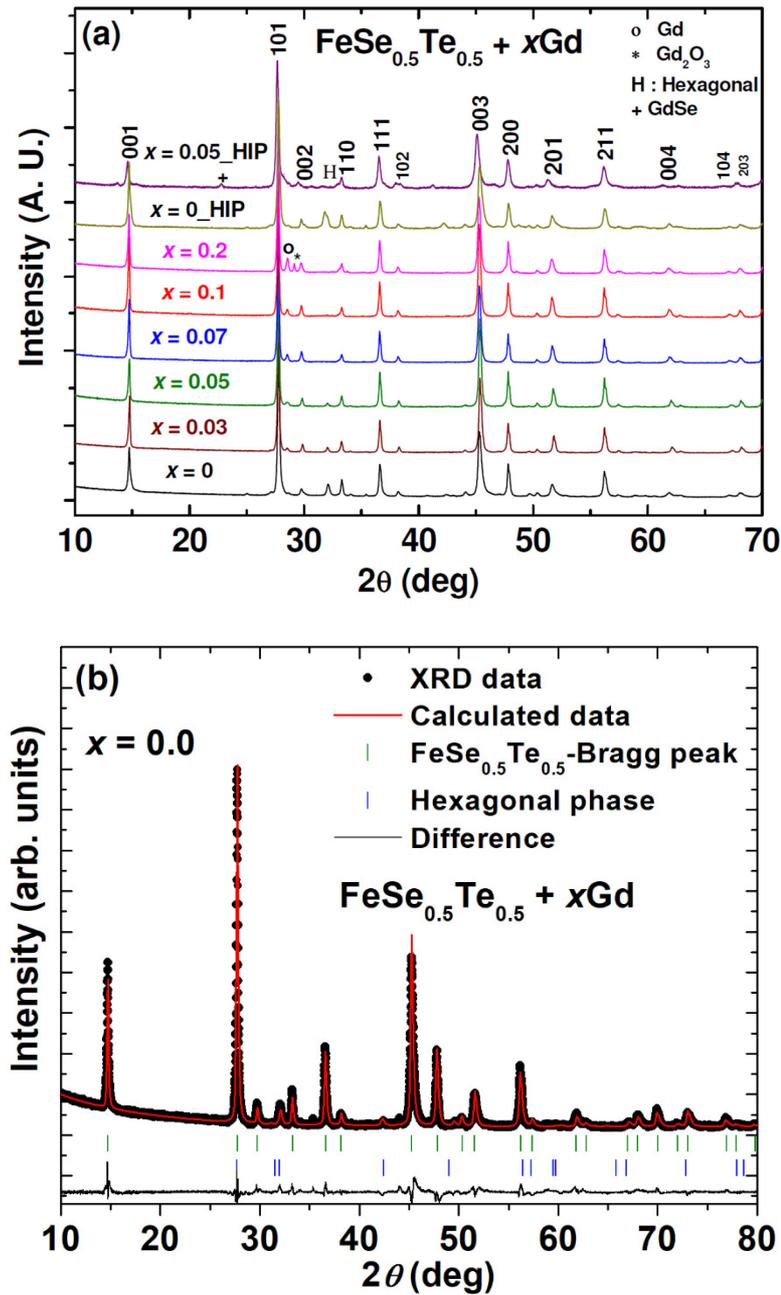

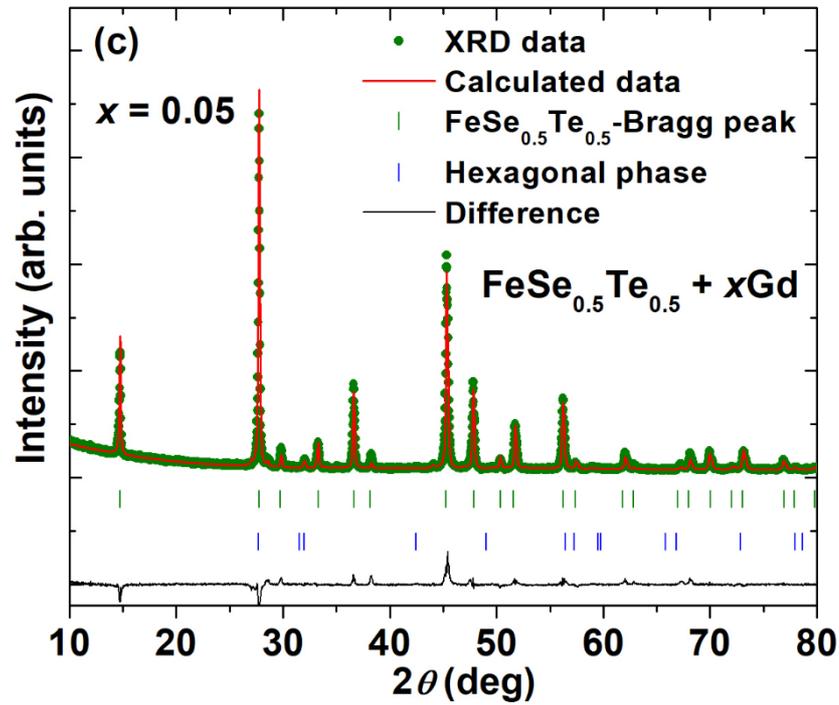

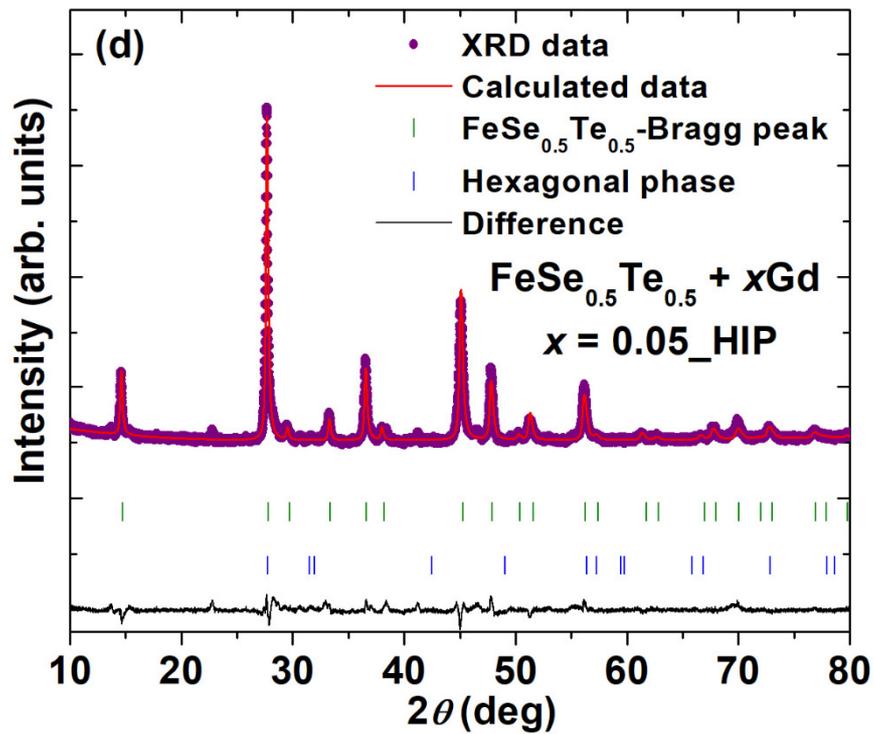

**Figure 2:** Mapping for the constituent elements of FeSe$_{0.5}$Te$_{0.5}$ + $x$Gd polycrystalline samples **(i)** the parent $x = 0$ **(ii)** $x = 0.05$ **(iii)** $x = 0.2$ prepared by CSP.

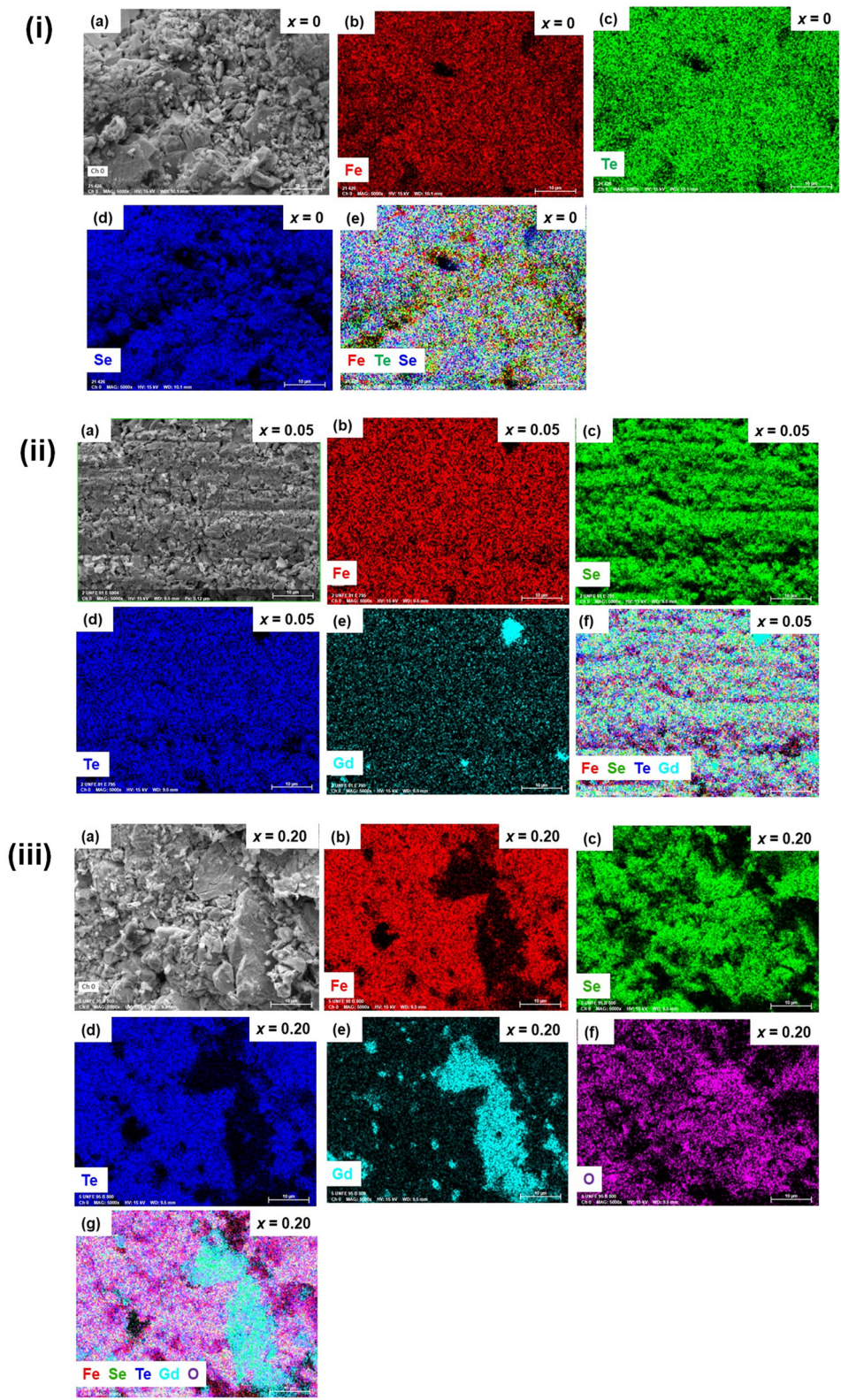

**Figure 3:** Mapping for the constituent elements of FeSe$_{0.5}$Te$_{0.5}$ + $x$Gd polycrystalline samples **(i)** the parent $x = 0$\_HIP **(ii)** $x = 0.05$\_HIP prepared by HP-HTS.

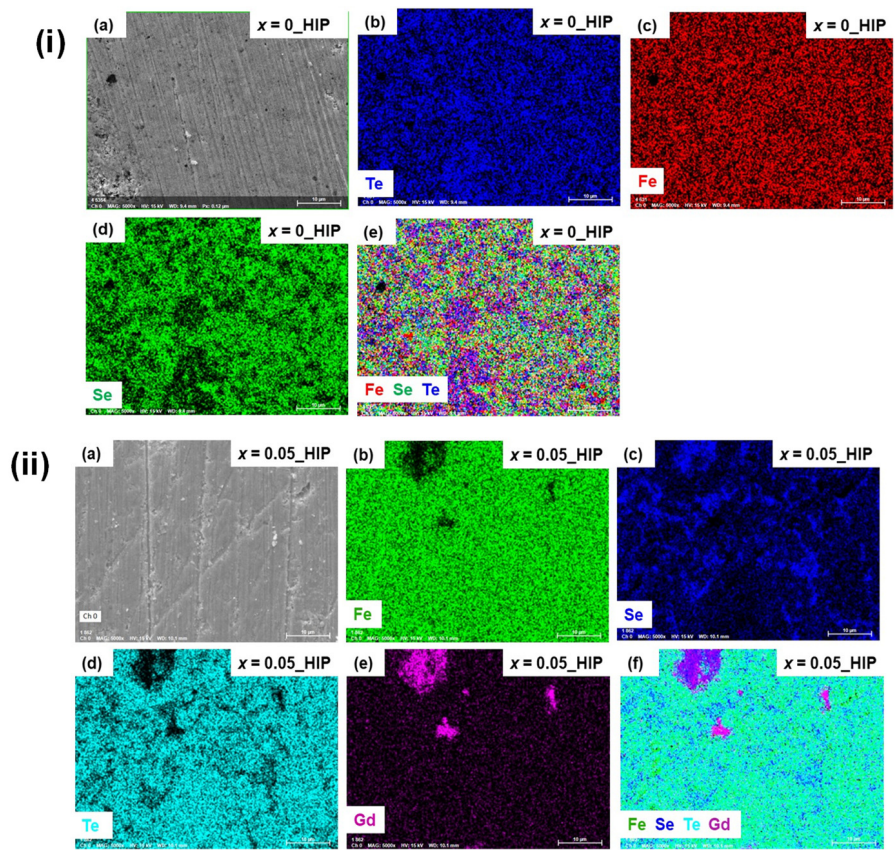

**Figure 4:** Back-scattered (BSE) images of FeSe$_{0.5}$Te$_{0.5}$ + $x$Gd polycrystalline samples: **(a)-(c)** for $x$ = 0; **(d)-(f)** for $x$ = 0.05; **(g)-(i)** for $x$ = 0.2. Light gray, bright and black contrast correspond to FeSe$_{0.5}$Te$_{0.5}$, Gd$_2$O$_3$, and pores, respectively

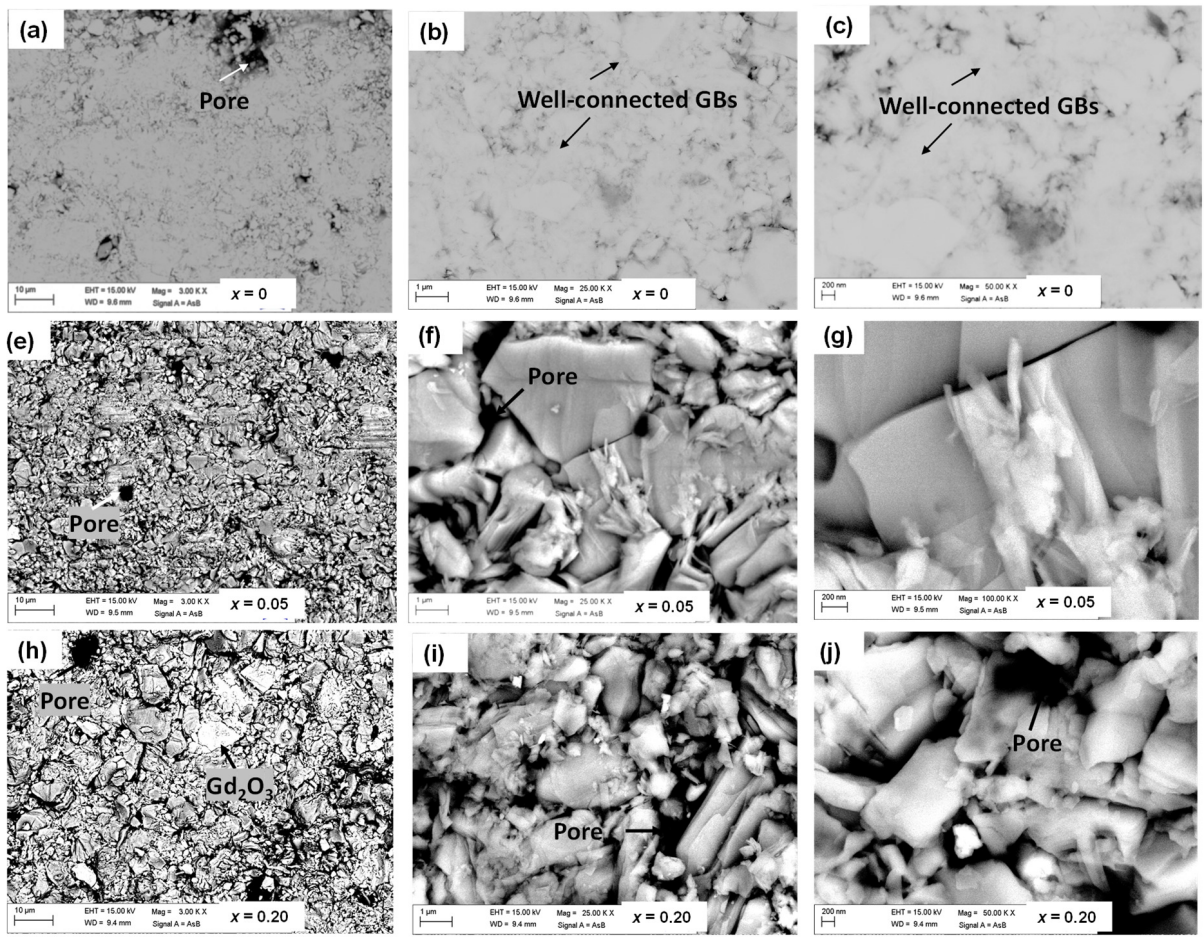

**Figure 5:** Back-scattered (BSE) images of FeSe$_{0.5}$Te$_{0.5}$ + $x$Gd polycrystalline samples: **(a)-(c)** for $x =$ 0_HIP; **(d)-(f)** for $x = 0.05$_HIP prepared by HP-HTS. Light gray, bright and black contrast correspond to FeSe$_{0.5}$Te$_{0.5}$, Gd$_2$O$_3$/GdSe, and pores, respectively.

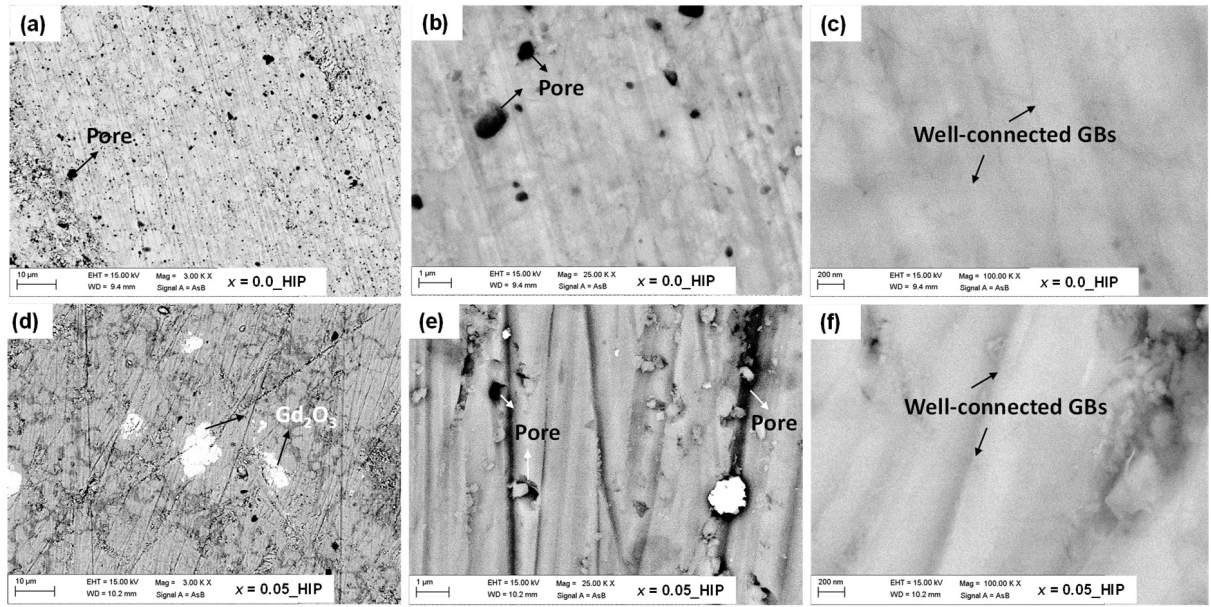

**Figure 6:** The temperature dependence of the magnetic susceptibility ($\chi = 4\pi M / H$) measured under zero field-cooled (ZFC) and field-cooled (FC) mode in an applied magnetic field $\mu_0 H = 20$ Oe for **(a)** FeSe$_{0.5}$Te$_{0.5}$ + $x$Gd ($x = 0$ and $x = 0.05$) bulks prepared by CSP and **(b)** $x = 0\_$HIP, $x = 0.05\_$HIP prepared by HP-HTS.

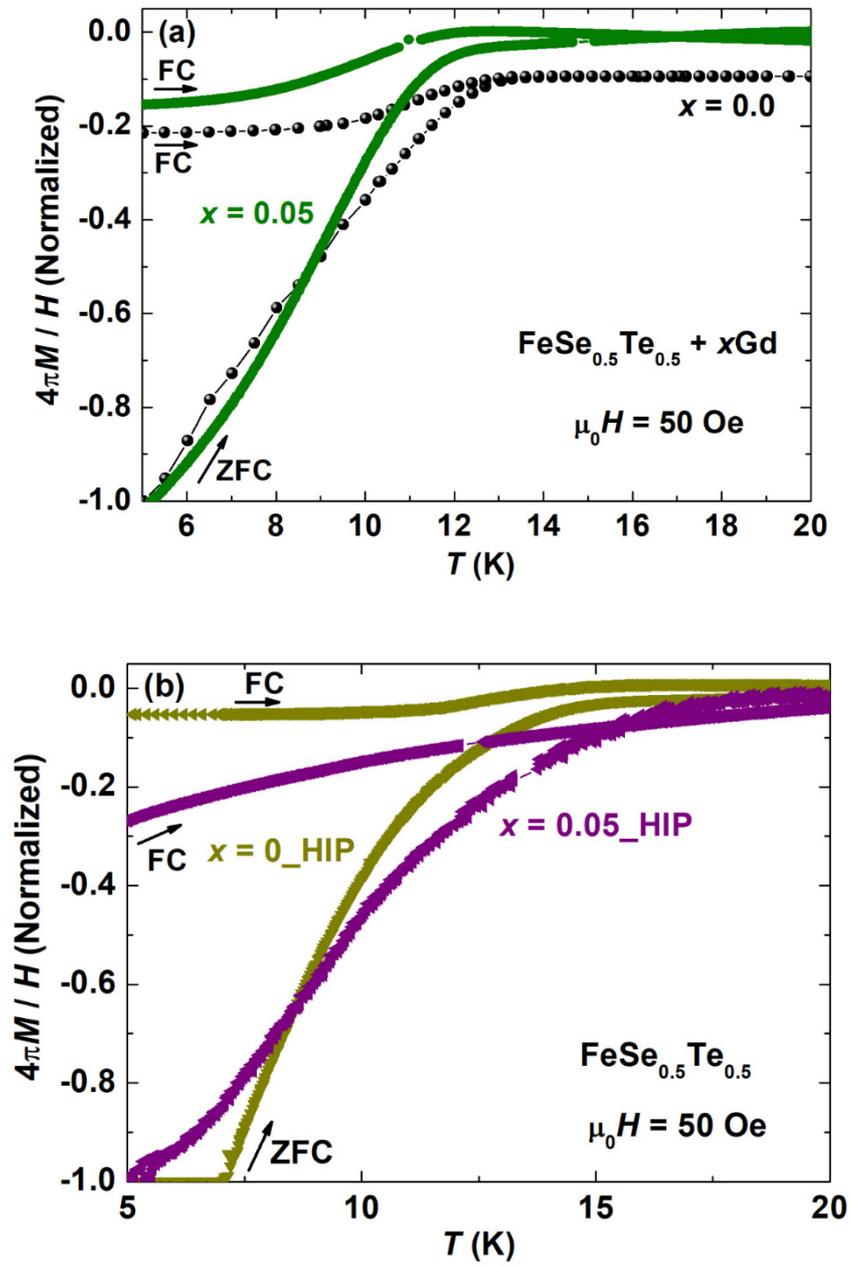

**Figure 7: (a)** The variation of resistivity ($\rho$) with the temperature for Gd added FeSe$_{0.5}$Te$_{0.5}$ (FeSe$_{0.5}$Te$_{0.5}$ + $x$Gd ($x$ = 0, 0.03, 0.05, 0.07, 0.10) prepared by CSP. The inset shows the resistivity variation of the sample $x$ = 0.2 in the whole temperature range. **(b)** Low temperature dependence of the resistivity behaviours for various samples prepared by CSP in low temperature region (9-20 K). **(c)** The temperature dependence of the low temperature resistivity for FeSe$_{0.5}$Te$_{0.5}$ + $x$Gd ($x$ = 0, 0.03, 0.05, 0.07, 0.10) with respect to different currents $I$ = 5, 10, 20 mA. **(d)** The temperature variation of resistivity in the whole temperature range and **(e)** Low temperature variation of resistivity for sample $x$ = 0.05_HIP and 0_HIP prepared by HP-HTS with the parent compound $x$ = 0 prepared by CSP.

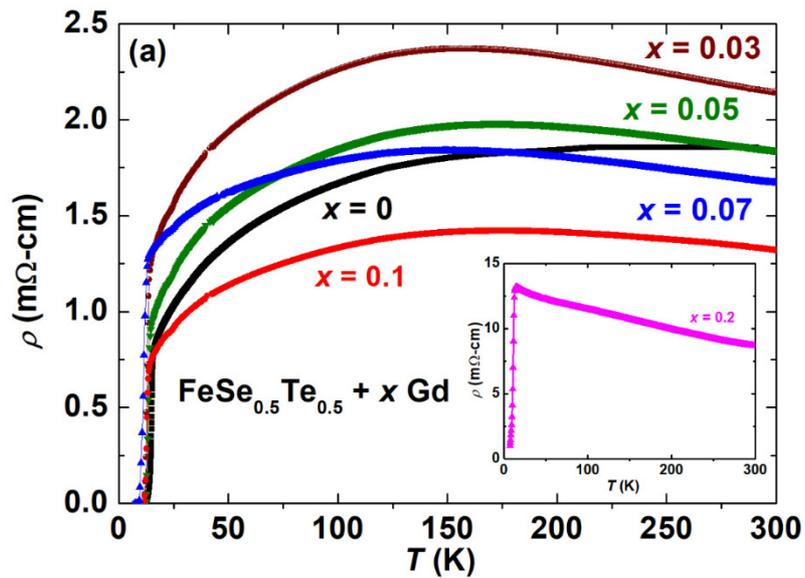

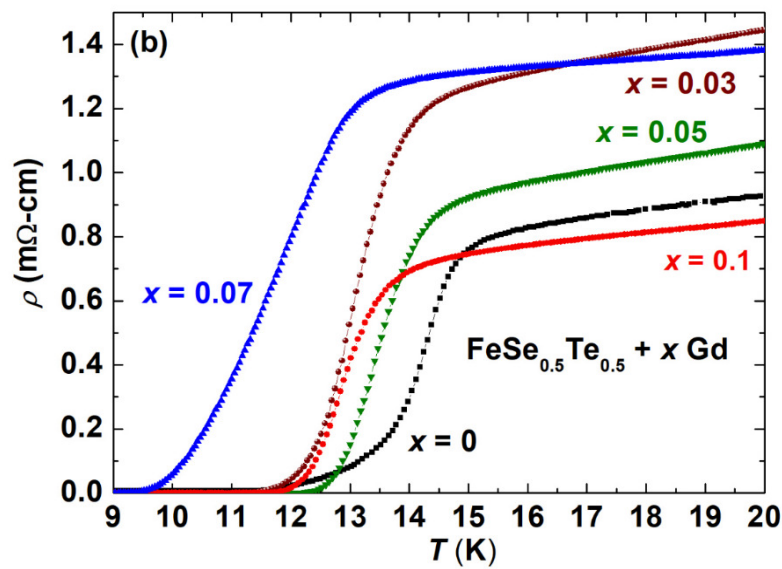

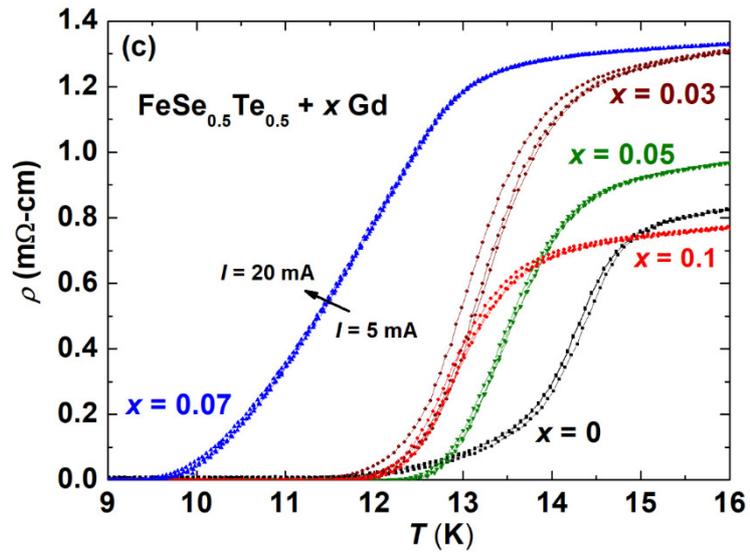

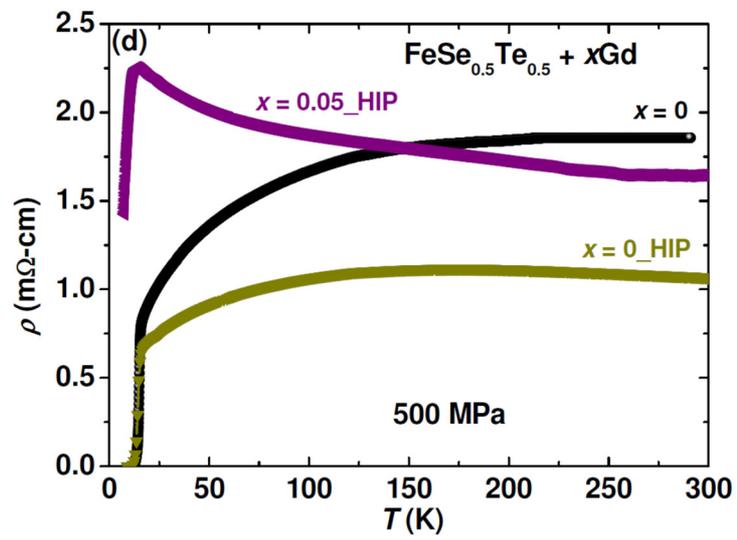

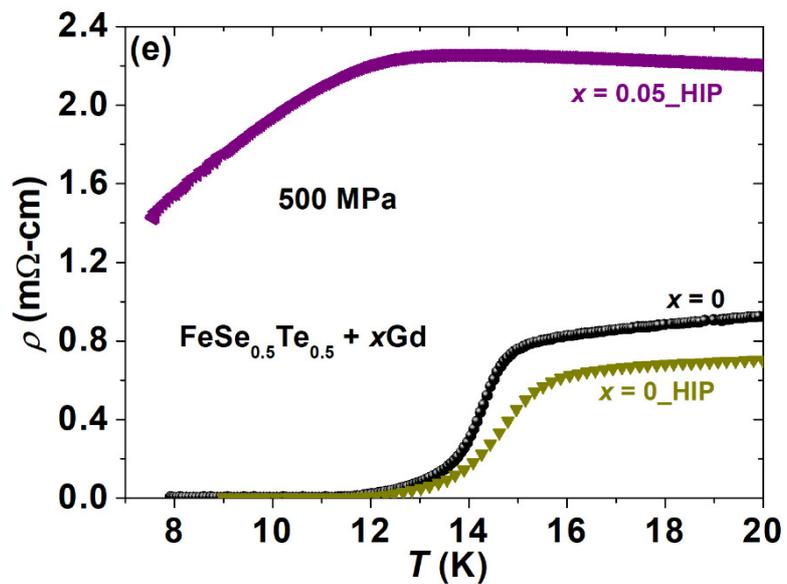

**Figure 8:** The critical current density ($J_c$) variation with the magnetic field ($H$) for $FeSe_{0.5}Te_{0.5} + x$Gd ($x = 0, 0.05, 0.05\_HIP$, and $x = 0\_HIP$) samples up to 9 T at temperature of 7 K. The inset figure shows the magnetic hysteresis loop $M(H)$ at 7 K for $x = 0.0\_HIP$ after the subtraction of the normal state background ($M(H)$ at 22 K).

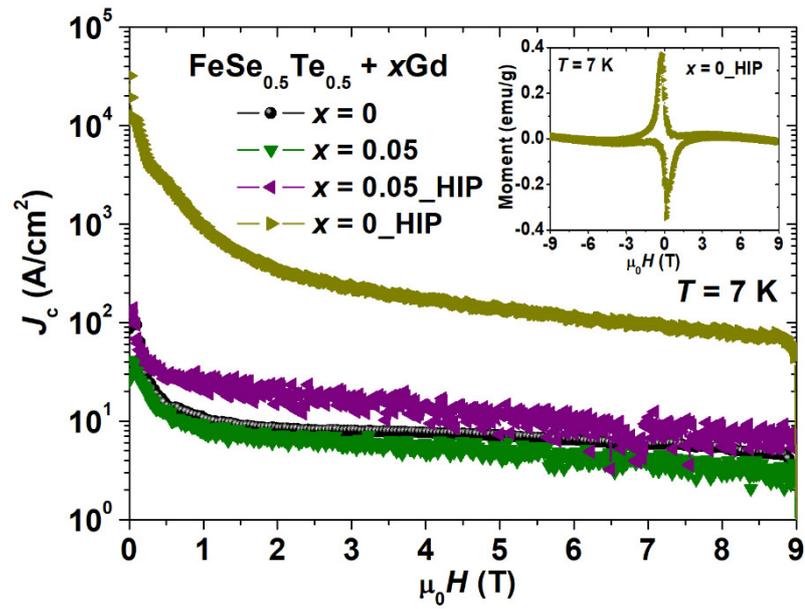

**Figure 9:** The variation of **(a)** lattice parameter '*c*' **(b)** the onset transition temperature ($T_c$), **(c)** transition width ($\Delta T$), **(d)** room temperature resistivity $\rho_{300K}$ and **(e)** residual resistivity ratio *RRR* ($\rho_{300K}$ / $\rho_{20K}$) with weight% of Gd addition for the parent FeSe$_{0.5}$Te$_{0.5}$ *i.e.* FeSe$_{0.5}$Te$_{0.5}$ + *x*Gd ($x = 0, 0.03, 0.05, 0.07, 0.1,$ and $0.2$) prepared by CSP and HP-HTS.

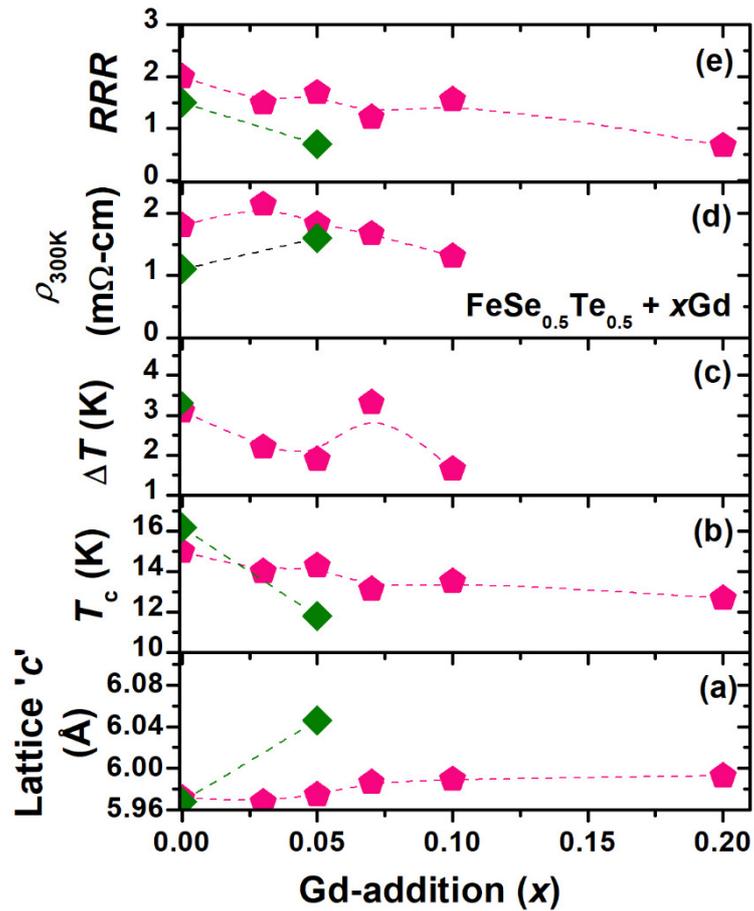